\documentclass[prb,twocolumn,showpacs]{revtex4}
\usepackage{amsmath}
\usepackage{graphicx}

\newcommand{\be}{\begin{equation}}
\newcommand{\ee}{\end{equation}}

\newcommand{\tb}{Tb$_2$Ti$_2$O$_7$}

\newcommand{\Journal}[4]{{\em #1} \textbf{#2}, #3 (#4)}
\newcommand{\PRev}{Phys.\ Rev. }

\begin{document}

\title{Quantum spin configurations in Tb$_2$Ti$_2$O$_7$}

\author{S. H. Curnoe}
\email[Electronic address: ]{curnoe@physics.mun.ca}
\affiliation{Department of Physics and Physical Oceanography,
Memorial University of Newfoundland, St.\ John's, Newfoundland \& Labrador 
A1B 3X7, Canada}

\begin{abstract}
Low energy collective angular momentum states of the Tb$^{3+}$ ions in
\tb\ are classified according to the irreducible representations of the 
octahedral point group. Degeneracy lifting due to the exchange interaction
is discussed.  Diffuse neutron scattering intensity patterns are
calculated for each collective angular momentum state and the 
ground state is inferred by comparing to experiment. 
\end{abstract}


\maketitle

%\section{Introduction}
\tb\ is geometrically frustrated due to its
pyrochlore crystal structure, apparently leading to a ``spin liquid"
state which persists to  extremely low temperatures.\cite{gardner2003}  
The spin liquid
%or ``collective paramagnetic"
is characterised by an abundance of low-lying excited states
which tend to prevent the formation of an ordered state
because of fluctuations.
The focus of experimental work has been to determine the conditions under
which the spin liquid state becomes unstable against the
development of long range order.  Theoretical studies
of antiferromagnetic interactions predict that ordering should occur
with a N\'{e}el temperature of approximately 1.2 K,\cite{gingras2000,hertog2000,kao2003}  but this is not seen
experimentally.  
There are reports of
short-range ordering occurring at $T_c=0.37$ K\cite{yasui2002,hamaguchi2004} 
and a spin glass phase  below 200 mK,
\cite{luo2001,hamaguchi2004,jana2003} 
however one group has found that the spin liquid
phase persists down to at least 50 mK.\cite{gardner2003,gardner1999}
%A combination of isotropic and
%uniaxial pressure produce an antiferromagnetically
%ordered state with wavevector
%$k=(1,0,0)$,\cite{mirabeau2004} while a magnetic field, whether combined 
%with uniaxial pressure or not, produces an ordered state with 
%wavevector $k=0$.\cite{yasui2001,mirabeau2004,rule2006}
Even in the absence of long range order, one may expect that local correlated
spin configurations, possibly governed by the exchange interaction, 
will be present.  Indeed, a diffuse neutron scattering experiment 
that found an intensity pattern characterised by a broad peak at 
${\bf q} = [0,0,2]$\cite{gardner2001} has been interpreted in this way.\cite{molavian}

There is a vast literature of theoretical works on the general topic 
of spin liquids in geometrically frustrated systems.\cite{anderson}
A small subset of these treat the magnetic ions
as classical Ising spins which point toward the corners of the
tetrahedra, and this approach has been useful for
understanding spin ice behaviour in the related compounds
Ho$_2$Ti$_2$O$_7$ and Dy$_2$Ti$_2$O$_7$.\cite{hertog2000,harris1997}
In this article, we extend this approach using 
basic considerations of symmetry and quantum mechanics.

\tb\ crystallises in the pyrochlore structure with space group 
$Fd\bar{3}m$ (\#227, O$_h^7$).
With two copies of the chemical formula per unit cell, there are four Tb ions
which occupy the corners of a corner-shared tetrahedral network.
The local site symmetry is $\bar{3}$m (D$_{3d}$), where the three-fold
axes point along the directions $[111]$, $[-1-11]$, $[-11-1]$ and
$[1-1-1]$ for 
sites \#1, 2, 3 and 4, respectively.  In this environment, the $J=6$  manifold of the 4f$^8$ state
of Tb$^{3+}$
splits into levels labeled by the irreducible representations (IR's) of D$_3$: 
$3A_1\oplus 2A_2 \oplus 4 E$.\cite{tinkham}   Neutron scattering 
measurements and {\em ab initio} calculations find that the
ground state and first excited states are $E$ doublets, separated by 1.5 meV.\cite{gingras2000}

Both the ground state and first excited state 
are linear combinations of the angular momentum states
$|\pm5\rangle$, $|\pm2\rangle$, $|\mp1\rangle$ and $|\mp4\rangle$,
where the quantum numbers measure $\vec{J}$ in the direction of the
$C_3$ axis.
The ground state doublet was determined to be\cite{gingras2000,note}
\begin{equation}
|\pm\rangle =  \pm 0.13 |\pm 5\rangle \mp 0.13 |\mp 1\rangle -.97 |\mp 4\rangle.
\label{doublet}
\end{equation}
With four Tb ions per unit cell, there is therefore a
sixteen-fold degeneracy of the ground state for each unit cell.
The collective
angular momentum states are denoted 
$|\pm\pm\pm\pm\rangle \equiv
 |\pm\rangle_1\otimes|\pm\rangle_2\otimes|\pm\rangle_3\otimes|\pm\rangle_4$,
where the subscripts indicate the site number.
These states are divided
according to the representations by which 
they transform under the operations of the octahedral point group
O$_h$, the underlying point group 
symmetry of the crystal, as 
$
A_{1g}\oplus3E_g\oplus2T_{1g}\oplus
T_{2g}$.\cite{note2}
The actual states are listed in Table I.
The subscript $g$ will be henceforth omitted
since we will always refer to even representations.
\begin{table*}
\begin{tabular}{rl}
\hline
$|A_1\rangle=$ & $(|++--\rangle + |+-+-\rangle +|+--+\rangle +|-++-\rangle +|-+-+\rangle
+|--++\rangle)/\sqrt{6}$ \\ \hline
$|E^{(1)}_{+}\rangle =$ & $
             |++++\rangle$ \\
  $|E^{(1)}_{-}\rangle=$ & $            |----\rangle $ \\ \hline
$|E^{(2)}_{+}\rangle = $ & $ 
       (|+---\rangle + |-+--\rangle + |--+-\rangle +|---+\rangle)/2$ \\
$|E^{(2)}_{-}\rangle =$ & $        |+++-\rangle + |++-+\rangle + |+-++\rangle +|-+++\rangle)/2 $ \\ \hline
$|E^{(3)}_{+}\rangle=$ & $ 
         (|++--\rangle + \varepsilon|+-+-\rangle + \varepsilon^2|+--+\rangle
          +|--++\rangle + \varepsilon|-+-+\rangle + \varepsilon^2|-++-\rangle)\sqrt{6}$\\
$|E^{(3)}_{-}\rangle=$ & $\mbox{c.c.}
$\\ \hline
$
|T_{1x}^{(1)}\rangle =$ & 
	$(\varepsilon^2[-|+++-\rangle + |++-+\rangle + |+-++\rangle-|-+++\rangle]
+\varepsilon[
         |+---\rangle - |-+--\rangle -|--+-\rangle +|---+\rangle])/2\sqrt{2}$\\
$|T_{1y}^{(1)}\rangle =$ & $  (\varepsilon[|+++-\rangle - |++-+\rangle + |+-++\rangle-|-+++\rangle]
         +\varepsilon^2[|+---\rangle - |-+--\rangle +|--+-\rangle -|---+\rangle])/2\sqrt{2}$\\
$|T_{1z}^{(1)}\rangle =$ & $(|+++-\rangle+|++-+\rangle - |+-++\rangle-|-+++\rangle
         +|+---\rangle + |-+--\rangle -|--+-\rangle -|---+\rangle)/2\sqrt{2}
$\\  \hline
$|T_{1x}^{(2)}\rangle =$ & $
        (|+--+\rangle - |-++-\rangle )/\sqrt{2}$\\

$|T_{1y}^{(2)}\rangle = $  & $ (|+-+-\rangle - |-+-+\rangle)/\sqrt{2}$\\  
$|T_{1z}^{(2)}  \rangle = $ & $
(|++--\rangle-|--++\rangle )/\sqrt{2} $\\ \hline

$
|T_{2x}\rangle = $ 
 & $ (\varepsilon^2[-|+++-\rangle + |++-+\rangle + |+-++\rangle-|-+++\rangle]
         -\varepsilon[|+---\rangle - |-+--\rangle -|--+-\rangle +|---+\rangle])/2\sqrt{2}$\\
$|T_{2y}\rangle =$ & 
        $(\varepsilon
[|+++-\rangle - |++-+\rangle + |+-++\rangle-|-+++\rangle]
         -\varepsilon^2[|+---\rangle - |-+--\rangle +|--+-\rangle -|---+\rangle])/2\sqrt{2}$\\
$|T_{2z}  \rangle = $ & $
(|+++-\rangle+|++-+\rangle - |+-++\rangle-|-+++\rangle
         -|+---\rangle - |-+--\rangle +|--+-\rangle +|---+\rangle)/2\sqrt{2}
$ \\ \hline
\end{tabular}
\caption{
Basis functions of the 
collective angular momentum states for the four Tb ion
sites, labeled according to the
irreducible representations of $O_h$  by which they transform. 
$\varepsilon = \exp(2\pi i/3)$.  }
\end{table*}

%\section{Antiferromagnetic Interaction}

We begin by considering 
the nearest neighbour antiferromagnetic (AF) exchange interaction. 
${\cal J} \sum_{<ab>}\vec{J}_a
\cdot\vec{J}_b
$
is isotropic, therefore it reduces the
degeneracy of the 16-fold collective angular momentum states
without mixing the different representations.
To calculate the matrix elements,
$\vec{J}_a$ must be expressed in terms of the angular momentum quantisation
axes (the $C_3$-axis of each Tb ion).  
In the following, $J_{ai}$ will always refer to the $i$-axis  ($i=x,y,z$)
using the local coordinate axes of the $a$ position ($a=1,2,3,4$), while
$J^i_a$ (used below, in the discussion of neutron scattering)
refers to global axes. 
For the specific choice of axes for the 1 and 2 positions,
\begin{eqnarray}
\hat{x}_1 = (1,1,-2)/\sqrt{6},  &&  \hat{x}_2 = (-1,-1,-2)/\sqrt{6} \nonumber\\
\hat{y}_1 = (-1,1,0)/\sqrt{2},  && \hat{y}_2 = (1,-1,0)/\sqrt{2} \nonumber\\
\hat{z}_1 = (1,1,1)/\sqrt{3}, && \hat{z}_2 = (-1,-1,1)/\sqrt{3} ,
\label{12}
\end{eqnarray}
one finds
\begin{eqnarray}
\vec{J}_1\cdot\vec{J}_2&=&-\frac{1}{3}J_{1z}J_{2z} \nonumber \\
& & - \frac{\sqrt{2}}{3}
[J_{1z}(J_{2+}+J_{2-})+(J_{1+}+J_{1-})J_{2z}] \nonumber \\
&& +\frac{1}{3}(J_{1+}J_{2+}+J_{1-}J_{2-}) \nonumber \\
& & -\frac{1}{6}(J_{1+}J_{2-}+J_{1-}J_{2+}).
\label{ex12}
\end{eqnarray}
For the axes of the 3 and 4 positions,
\begin{eqnarray}
\hat{x}_3 = (-1,1,2)/\sqrt{6},  &&  \hat{x}_4 = (1,-1,2)/\sqrt{6} \nonumber\\
\hat{y}_3 = (1,1,0)/\sqrt{2},  && \hat{y}_4 = (-1,-1,0)/\sqrt{2} \nonumber\\
\hat{z}_3 = (-1,1,-1)/\sqrt{3}, && \hat{z}_4 = (1,-1,-1)/\sqrt{3} 
\label{34}
\end{eqnarray}
some non-trivial phases appear:
\begin{eqnarray}
\vec{J}_1\cdot\vec{J}_3&=&-\frac{1}{3}J_{1z}J_{3z} \nonumber \\
& & - \frac{\sqrt{2}}{3}
[J_{1z}(\varepsilon J_{3+}+\varepsilon^2 J_{3-})+(\varepsilon J_{1+}+\varepsilon^2 J_{1-})J_{3z}] \nonumber \\
&& +\frac{1}{3}(\varepsilon^2 J_{1+}J_{3+}+\varepsilon J_{1-}J_{3-}) \nonumber \\
& & -\frac{1}{6}(J_{1+}J_{3-}+J_{1-}J_{3+}) \label{ex13}\\
\vec{J}_1\cdot\vec{J}_4&=&-\frac{1}{3}J_{1z}J_{4z} \nonumber \\
& & - \frac{\sqrt{2}}{3}
[J_{1z}(\varepsilon^2 J_{4+}+\varepsilon J_{4-})+(\varepsilon^2 J_{1+}+\varepsilon J_{1-})J_{4z}] \nonumber \\
&& +\frac{1}{3}(\varepsilon J_{1+}J_{4+}+\varepsilon^2 J_{1-}J_{4-}) \nonumber \\
& & -\frac{1}{6}(J_{1+}J_{4-}+J_{1-}J_{4+}) \label{ex14}
\end{eqnarray}
where $\varepsilon = \exp(2 \pi i /3)$.
The expression for $\vec{J}_2\cdot\vec{J}_3$ is similar to $\vec{J}_1\cdot\vec{J}_4$,
etc.
%$\vec{J}_2\cdot\vec{J}_4$ is similar to $\vec{J}_1\cdot\vec{J}_3$ and
%$\vec{J}_3\cdot\vec{J}_4$ is similar to $\vec{J}_1\cdot\vec{J}_2$.
The matrix elements are 
\begin{eqnarray}
\langle \pm | J_z | \pm \rangle &\equiv& \pm j \\
\langle \pm | J_{\pm} | \mp \rangle &\equiv& t \label{t}
\end{eqnarray}
For the states given in Eq.\ \ref{doublet}, $j=-3.69\hbar$ and $t$ is exactly
zero 
(this is why the spins are Ising-like).
The $E^{(1)}$ states have the lowest energy $-2{\cal J}j^2$, 
while the $E^{(2)}$, $T_1^{(1)}$ and $T_2$ states
have zero energy, and the $A_1$, $E^{(3)}$ and $T_1^{(2)}$ states have energy
$2{\cal J}j^2/3$.  
Note that any linear combination of $A_1$, $E^{(3)}$ and $T_1^{(2)}$ 
resembles the ``spin ice" state (two spins pointing in and 
two spins pointing out of each tetrahedron), 
which is the ground state for ferromagnetic exchange on 
classical spins.  The $E_{\pm}^{(1)}$ states resemble the 
classical states often described in the literature as the spins
pointing outward from the corners of the cube which contains the tetrahedron,
which also minimise the AF exchange energy.  

Because $t=0$, the exchange interaction leads to
no mixing between states listed in Table I 
belonging to the same representation,
although this would be permitted under general
considerations of symmetry. 
Non-zero matrix elements that mix $|+\rangle$ and $|-\rangle$ states
 could originate from
mixing with higher crystal field
levels\cite{molavian} or with higher multipole interactions, beginning
with quadrupole.  Obviously such effects are
relevant if the exchange interaction constant ${\cal J}$ is small.
If $t\neq 0$ then the mixing between states due to the exchange interaction
is easily found using the matrix elements given in Table III, below.
The dipole-dipole interaction, when limited to a single
tetrahedron, only renormalises the exchange interaction, 
with slight differences between the diagonal and off-diagonal 
matrix elements of the sixteen collective angular momentum states. 

To determine which of the sixteen states is the most likely ground state
for \tb\ we compare the 
diffuse neutron scattering intensity pattern with 
calculations of the scattering function.
In the dipole approximation, the scattering intensity is proportional
to\cite{jensen} 
\begin{equation}
I({\bf q}) \propto e^{-2 W(q)}|F({\bf q})|^2
\sum_{m}e^{-E_m/k_BT}f_m(q)
\end{equation}
where $W(q)$ is the Debye-Waller factor, $F({\bf q})$ is the form factor
for Tb$^{3+}$ and
\begin{equation}
f_m({\bf q}) = \sum_{i,j}\sum_{a,b}
\sum_{n}
(\delta^{ij}-\hat{q}^i\hat{q}^j)
\langle m|J_a^i|n\rangle\langle n|J_b^j|m\rangle e^{i{\bf q}\cdot ({\bf r}_b
-{\bf r}_a)}.
\label{fm(q)}
\end{equation}
$a$ and $b$ are the four sites at the corners of the tetrahedron and
$n$ and $m$ are the sixteen states, whose energy differences are assumed
to
lie within the energy range across which the neutron scattering 
is integrated.
The functions $f_m({\bf q})$ were calculated for each state $m$ and 
the results for states belonging to the same doublet or triplet 
were added.
The cases $j\neq 0$ and $t\neq0$ were considered separately.
The calculations yield four different patterns, shown in Fig.\ 1.
The correspondence between different patterns and states
is summarised in Table \ref{patterns}.

\begin{figure*}[htb]
\resizebox{!}{16cm}{\includegraphics{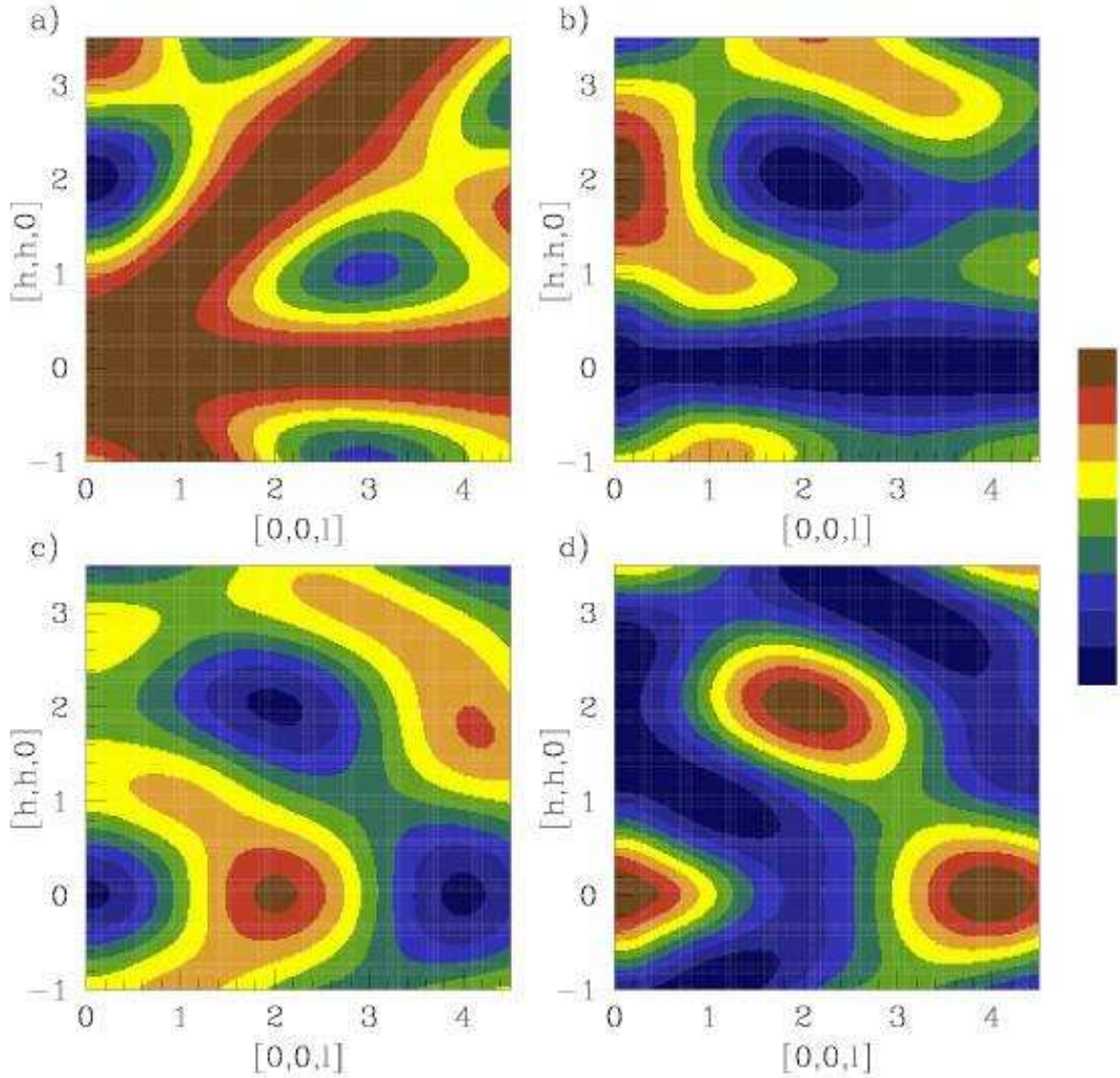}}
\caption{\label{patt}(Colour) Contour plots of
diffuse neutron intensity patterns $f_m({\bf q})$ (Eq.\ \ref{fm(q)})
for different states for non-zero $j$ (figure a) and non-zero $t$
(figures b-d)  (refer to Table II.)  In all four figures the intensities have
been scaled in order to draw them using the same colour range.}
\end{figure*}

\begin{table}[htb]
\begin{tabular}{|l|r|r|}
\hline
state & $j$ & $t$ \\ \hline
$A_1$ & a) & b) \\
$E^{(1)}$ & $-$a) & \\
$E^{(2)}$ & & b) \\
$E^{(3)}$ & a) & $-$b)\\
$T_1^{(1)}$ & & c) \\
$T_1^{(2)}$ & a) & \\
$T_2$ & & d) \\\hline
\end{tabular}
\caption{\label{patterns} The first column lists the states given in 
Table I, and the second and third columns indicate to which pattern of 
Fig.\ 1 each state corresponds, for non-zero $j$ ($t=0$) and non-zero $t$ 
($j=0$) respectively.  A
negative sign indicates that the intensity is reversed, and no entry indicates
that the pattern has a uniform intensity.}
\end{table}

If the exchange interaction plays a dominant role in this system
then one would expect that Fig.\ 1a), which 
is associated with 
the anti-ferromagnetic state $E^{(1)}$ and
the ferromagnetic states $A_1$, $E^{(3)}$ and $T_1^{(2)}$ for $j\neq 0$, 
would match diffuse neutron scattering measurements. 
However, this is clearly not the case.   Neutron scattering
patterns oscillate in three directions, $[0,0,l]$, $[0,l,l]$ and
$[l,l,l]$, and show a broad peak at the position
$[0,0,2]$.\cite{gardner2001}
The only figure which shows a qualitative agreement to experiment is
Fig.\ 1c), and in fact the agreement is very good.  
Considering that the experimental results are a sum over all $f_m({\bf q})$,
weighted by the factor $\exp(-E_m/k_B T)$, we conclude that the configuration
leading to Fig.\ 1c) is the ground state.

Fig.\ 1c) comes from
the state $T_1^{(1)}$,  which  is  neither ferro-magnetic nor
anti-ferromagnetic and 
completely
insensitive to the parameter $j$.
This result is difficult to explain, but at least it is
self-consistent.
The interaction which could lead to a $T_1^{(1)}$ type ground state
is unknown, but obviously it must contain an effective
non-zero $t$.  Likewise, the neutron scattering
pattern shown in Fig.\ 1c)  can only be derived with a non-zero $t$.
Moreover, a non-zero $t$ restores the true
quantum mechanical nature of this system, for without it only classical
Ising-like spins remain.\cite{kao2003,enjalran2004}

The origin of non-zero $t$ is still an open question but there are a couple of
possibilities worth considering.  First, higher order
multipole interactions restore quantum effects since the
matrix elements
\begin{equation}
\langle-|J_zJ_{-}|+\rangle = \langle-|J_{-}J_z|+\rangle  \label{mix1}
=-5.21 \hbar^2
\end{equation}
and
\begin{equation}
\langle-|J_{+}^2|+\rangle     \label{mix2} = -0.71 \hbar^2
\end{equation}
are non-zero.
Alternatively,
extending the  16-dimensional manifold
to include higher crystal electric field levels will add some 
non-zero matrix elements for the $J_{\pm}$ operators.  This approach was 
followed in Ref.\ \onlinecite{molavian} by including the first
excited crystal electric field doublet, thus increasing the
number of collective angular momentum states from 16 to 256.  In that case,
the calculated diffuse neutron scattering intensity pattern agreed well with
experiment but
a singlet ground state was found based on exchange and dipole-dipole
interactions.

The effect of a non-zero $t$ on the exchange interaction on
the 16 states given in Table I can be
calculated using (\ref{ex12}) (\ref{ex13}) and
(\ref{ex14}).  The results are summarised in Table \ref{ta_exchange}.
The lowest energy state is found in the $E$ sector for any 
values of $j$ and $t$.
Therefore, introducing a non-zero $t$ into the exchange interaction
rearranges the energy levels but never produces a
$T_1$ ground state.
\begin{table}[htb]
\begin{tabular}{|c|c|c|c|}
\hline
Sector & Exchange matrix & $t=0$ & $j=0$\\
\hline
$A_1$ & $2j^2/3-2t^2/3$ & $2j^2/3$ & $-2t^2/3$\\ \hline
$E$ & $\left(\begin{array}{ccc}
-2j^2 & 0 & t^2\sqrt{2/3}\\
0 & -t^2/2 & -4tj/\sqrt{3}\\
t^2\sqrt{2/3} & -4tj/\sqrt{3} & 2j^2/3+t^2/3\end{array}\right)$ 
& $\begin{array}{c} 2j^2/3 \\ 0 \\-2 j^2 \end{array}$ &
$\begin{array}{c} t^2 \\-t^2/2\\-2t^2/3\end{array}$ \\ \hline
$T_1$ & $\left(\begin{array}{cc}
-t^2/2 & 0 \\
0 & 2j^2/3\end{array}\right)$ & $\begin{array}{c} 2j^2/3\\0\end{array}
$ & $\begin{array}{c} 0 \\ -t^2/2 \end{array}$ \\  \hline
$T_2$ & $5t^2/6$ & 0 & $ 5t^2/6$ \\ \hline
\end{tabular}
\caption{\label{ta_exchange}Matrix elements for the exchange interaction
for each representation within the 16-state manifold spanned by the 
states in Table I.  The last two columns give the eigenvalues of the matrices
for $t=0$ and $j=0$ respectively.} 
\end{table}

To summarise, we have classified the low
energy collective angular momentum states of the
Tb$^{3+}$ ions according to the irreducible representations of the 
octahedral point
group $O_h$.  Diffuse neutron scattering patterns are best described by
the $T_1^{(1)}$ ground state.  This result cannot be accounted for by the exchange 
interaction.
Then, two important questions remain, namely, what is the nature of
the local interactions leading to such a ground state, and what are the 
low-lying excitations responsible for spin liquid behaviour.

%\section{Summary}

\begin{acknowledgments}
I thank Michel Gingras, Martin Plumer and 
Ivan Sergienko for valuable discussions.  This work was supported by 
NSERC of Canada.
\end{acknowledgments}


\begin{thebibliography}{00}
\bibitem{note} As written, this is imprecise, because the
$|\pm 4\rangle$ state has contributions from both the
$^7F_6$ (Hund's rules) manifold and the $^5G_6$ manifold.\cite{gingras2000}
\end{thebibliography}
\end{document}